\documentclass[aps,floatfix,showpacs,amsmath,amssymb,prb,superscriptaddress,12pt]{revtex4-1}
\usepackage{graphicx}
\usepackage{color}
\usepackage{ulem}
\usepackage{bm}
\usepackage{hyperref}
\usepackage{mathptmx}
\usepackage{soul}

\newcommand{\Cuso}{CuSO$_4\cdot$5D$_2$O}
\newcommand{\CusoH}{CuSO$_4\cdot$5H$_2$O}

\newcommand{\NL}{\newline}

\def\etal{\textit{et al.}}

\def\NL{$\newline$}
\def\NI{\noindent}

\def\anp{Ann.\ Phys.\ }

\def\jpcm{J.\ Phys.:\ Condens.\ Matter }

\def\jpsj{J.\ Phys.\ Soc.\ Jpn.\ }
\def\jsm{J.\ Stat.\ Mech.:\ Theory\ Exp.\ }
\def\nat{Nature (London) }
\def\natmat{Nature Mater.\ }
\def\natphys{Nature Physics }

\def\pla{Phys.\ Lett.\ A }
\def\prb{Phys.\ Rev.\ B }
\def\prl{Phys.\ Rev.\ Lett.\ }
\def\pr{Phys.\ Rev.\ }

\def\rmp{Rev.\ Mod.\ Phys.\ }
\def\rpp{Rep.\ Progr.\ Phys.\ }
\def\science{Science\ }
\def\ssc{Solid State Commun.\ }

\def\zpa{Z.\ Phys.\ A }

\begin{document}
\title{Fractional spinon excitations in the quantum Heisenberg \\ antiferromagnetic chain}
\author{Martin Mourigal}   
\affiliation{Institut Laue-Langevin, BP156, 38042 Grenoble Cedex 9,
                         France}
\affiliation{Laboratory for Quantum Magnetism, \'Ecole Polytechnique
             F\'ed\'erale de Lausanne (EPFL), 1015 Lausanne,
             Switzerland}      
\affiliation{Institute for Quantum Matter 
             and Department of Physics and Astronomy,
             Johns Hopkins University, Baltimore, MD 21218, USA}      
\author{Mechthild Enderle}
\affiliation{Institut Laue-Langevin, BP156, 38042 Grenoble Cedex 9,
                         France}
\author{Axel Kl\"{o}pperpieper}
\affiliation{Technische Physik, Universit\"{a}t des Saarlandes,
D-66041 Saarbr\"{u}cken, Germany}
\author{Jean-S\'ebastien Caux}
\affiliation{Institute for Theoretical Physics, University of Amsterdam, Science Park 904, 1090 GL Amsterdam, The Netherlands}
\author{Anne Stunault}
\affiliation{Institut Laue-Langevin, BP156, 38042 Grenoble Cedex 9,
                         France}
\author{Henrik M. R{\o}nnow}
\affiliation{Laboratory for Quantum Magnetism, \'Ecole Polytechnique
             F\'ed\'erale de Lausanne (EPFL), 1015 Lausanne,
             Switzerland}
\date{\today}
\maketitle

{\bf Assemblies of interacting quantum particles often surprise us with properties that are difficult to predict. One of the simplest quantum many-body systems is the spin 1/2 Heisenberg antiferromagnetic chain, a linear array of interacting magnetic moments. Its exact ground state is a macroscopic singlet entangling all spins in the chain. Its elementary excitations, called spinons, are fractional spin 1/2 quasiparticles; they are created and detected in pairs by neutron scattering.  Theoretical predictions show that two-spinon states exhaust only 71\% of the spectral weight while higher-order spinon states, yet to be experimentally located, are predicted to participate in the remaining. Here, by accurate absolute normalization of our inelastic neutron scattering data on a compound realizing the model, we account for the full spectral weight to within 99(8)\%. Our data thus establish and quantify the existence of higher-order spinon states. The observation that within error bars, the entire weight is confined within the boundaries of the two-spinon continuum, and that the lineshape resembles a rescaled two-spinon one, allow us to develop a simple physical picture for understanding multi-spinon excitations.}


100 years ago Max von Laue and co-workers discovered X-ray diffraction~\cite{Friedrich12}, thereby giving birth to the field of crystallography to which we owe much of our understanding of materials on the atomic scale. The very first diffraction image was recorded from a single crystal of copper sulphate pentahydrate~\cite{Friedrich12,Friedrich52}. In addition to vast practical use including herbicide, wood impregnation and algae control in swimming pools, copper sulphate also carries great educational importance. Generations of school children have been inspired in chemistry classes across the globe by growing from evaporating solution beautiful blue crystals of copper sulphate (in 2008, artist Roger Hiorns created an installation called {\it Seizure}~\cite{Seizure08} covering an entire apartment in copper sulphate crystals). When cooled close to absolute zero temperature copper sulphate has even more fascinating lessons to teach -- it becomes a quantum spin liquid. Moreover, it materializes one of the simplest models hosting complex quantum many body physics, the one-dimensional spin 1/2 Heisenberg antiferromagnet, for which there exists an exact analytic solution -- namely the Bethe ansatz~\cite{Bethe31}. 

Quantum spin liquid ground states entangle a macroscopic number of spins and give rise to astonishing and counter-intuitive phenomena. Quantum spin liquids occur in a variety of contexts ranging from the quantum spin Hall effect~\cite{Raghu08,Kane05} over high-T$_c$ superconductivity \cite{Anderson87,Kivelson87,Lee08} and confined ultracold gases to carbon nanotubes \cite{Meng10}. A particularly clear form of a gapless algebraic quantum spin liquid is realized in a one-dimensional array of spins 1/2 that are coupled by nearest-neighbour isotropic exchange, the spin 1/2 Heisenberg antiferromagnetic (HAF) chain. At zero temperature, this spin liquid is critical with respect to long-range antiferromagnetic order as well as with respect to dimerization \cite{Lieb61,Affleck89}. Its emerging gapless fractionalized excitations are called spinons~\cite{Fadeev81}. The concept of fractional excitations has been applied to magnetic monopoles in spin ice \cite{Castelnovo08,Morris09,Fennell09,Kadowaki09}, kagom\'{e} and hyper-kagom\'{e} lattices \cite{Hao09}, the quantum Hall effect~\cite{Laughlin83,Wilczek82,Halperin84,Kukushkin09}, conducting polymers \cite{Polymers,Polymers2}, and even to certain spin arrays with local spin larger than 1/2 \cite{Greiter09,Yao09}. For the prototypical spin 1/2 HAF chain, exact calculations of the dynamic structure factor over the whole range of the spectrum have become available. They reveal that two-spinon states exhaust 71\% of the first frequency moment sum-rule~\cite{Karbach97}. Including four-spinon states brings that level to 98(1)\%~\cite{Caux06}. The qualitative characteristics of two-spinon excitations, a continuum-like spectrum with linearly dispersing low-energy onset, are evidenced by inelastic neutron scattering on numerous compounds \cite{Heilmann78,Nagler91,Tennant93,Arai96,Dender96,Zheludev00,Coldea01,Stone03,Zaliznyak04,Lake05,Hong09,Thielemann09,Walters09,Lake10,Enderle10}. Among them, there are various quantitative attempts of an absolute comparison to theory~\cite{Nagler91,Zaliznyak04,Walters09}. However, none was sufficiently accurate to distinguish between an excitation continuum made of only two-spinon states and that composed of two- and higher-order spinon states, bearing a $\sim$30$\%$ larger spectral weight. The main sources of uncertainty come from the need to normalize the neutron scattering intensity to that of a reference Vanadium standard and from the role of other intrinsic and extrinsic sources of bias such as covalency, self-absorption, atomic zero point and thermal motion, and the limits of the dipole approximation used in the interpretation of the neutron cross section. The recent most accurate study, on the cuprate compound Sr$_2$CuO$_3$, finds  only 80\% of the predicted spectral intensity \cite{Walters09}.


Here we present a totally different approach to quantifying the full correlator, including two- and higher-spinon contributions. A magnetic field ${\bm H}||{\bm z}$ large compared to the antiferromagnetic exchange aligns all spins parallel. This fully polarized state, $\langle S_n^z\rangle=S$, is an eigenstate of the Heisenberg Hamiltonian $\mathcal{H}=J\,{\sum_n}{\bm S}_{n} {\bm S}_{n+1}-g\mu_B H S^z$, and identical to the classical ground state, which we obtain if we neglect all commutation relations of spin operators. In consequence, dispersion and intensity of the low-energy excitation spectrum are correctly described by linear spin-wave theory for which the elementary quasiparticles are non-interacting magnons. The classical magnon dispersion in the fully polarized phase has already been successfully employed to determine the microscopic parameters of the Hamiltonian of a two-dimensional frustrated quantum antiferromagnet \cite{Coldea02}. Here, we go one step further and determine not only the microscopic parameters from the dispersion of the magnon, but also exploit its wave-vector independent intensity to obtain an absolute intensity scale. Having fixed energy and intensity scale at high magnetic field, in the fully polarized ``classical'' phase, the quantum theory is tested against the zero-field data without any adjustable parameters. This approach avoids numerous uncertainties of previous attempts and allows to verify the role of higher-order spinon states quantitatively.

\begin{figure}[h!]
\includegraphics[width=0.93\columnwidth]{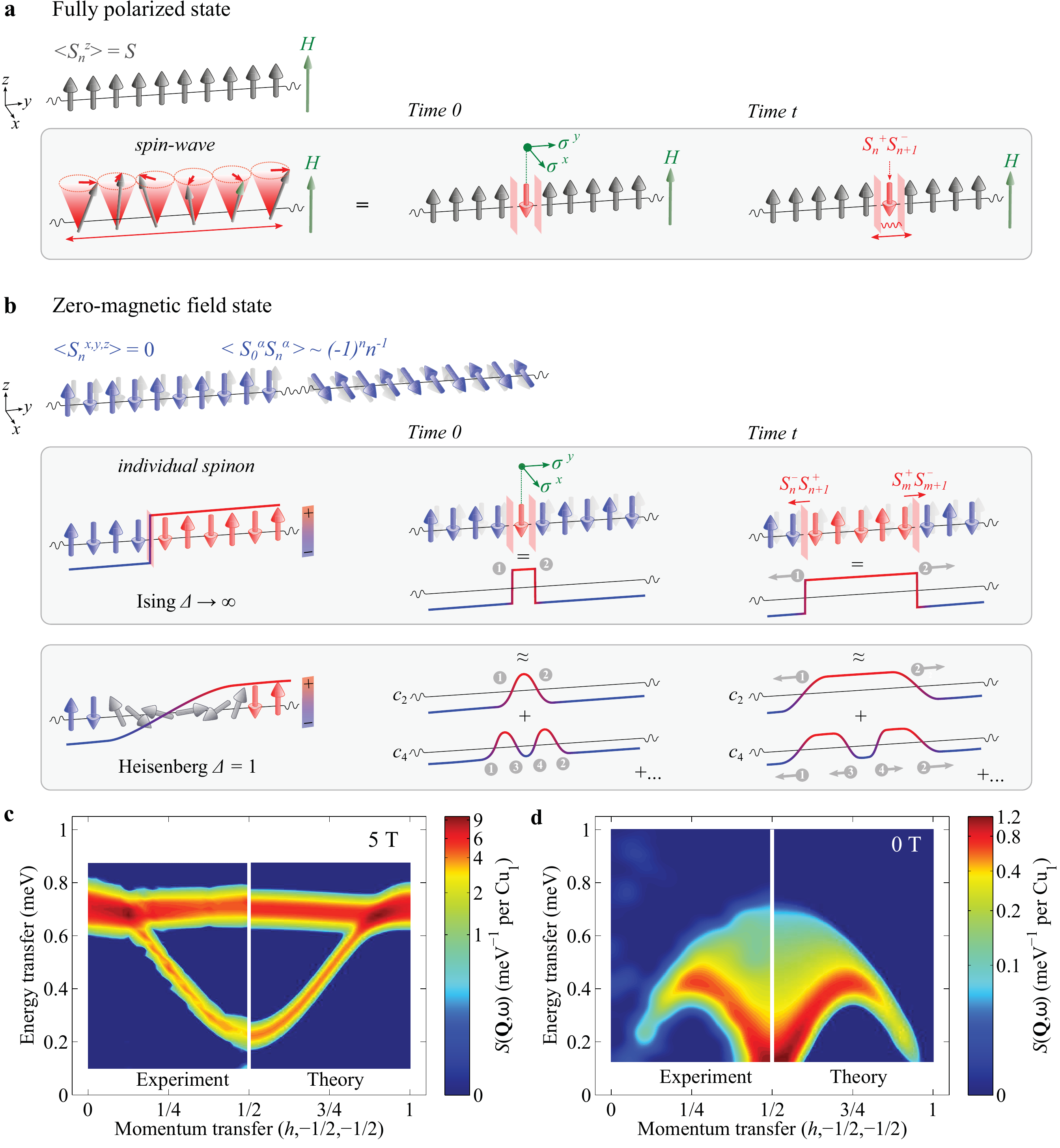}\\
\caption{}
\label{fig1}
\end{figure}

In Fig.~\ref{fig1}, we illustrate in a cartoon-like fashion the characteristic differences of the ground state and the excitations in the zero-field and the fully polarized phase. The magnon in the fully polarized state can be understood as a firmly bound domain wall pair that propagates and delocalizes as a single entity, Fig.~\ref{fig1}a. This results in a discrete energy-momentum dispersion relation of the magnon, evidenced in the intensity-colorplots Fig.~\ref{fig1}c. Domain wall propagation is achieved via the terms $S^{\alpha}_{n}S^{\alpha}_{n+1}$ in $\cal{H}$ with $\alpha=x,y$. The excitation amplitude (the magnon eigenvector) is therefore always transverse to the applied magnetic field ${\bm H}||{\bm z}$. 

In zero magnetic field the spins 1/2 entangle into a macroscopic singlet ${\bm S}_{\rm tot}={\sum_n}{\bm S}_n=0$, where local spin projections are no longer good quantum numbers, $\langle S_n^{\alpha}\rangle=0$, $\alpha=x,y,z$. Nevertheless, at $T=0$ the two-spin correlations decay only algebraically, $\langle S_0^{\alpha} S_{n}^{\alpha}\rangle \propto (-1)^n n^{-1}$, indicating infinitely large correlated antiferromagnetic regions. Snapshots of such a correlated region are shown in Fig.~\ref{fig1}b. Within such a region, the inelastically scattered neutron provokes $S_{\rm tot}=1$ excitations which we may first imagine as a local spin-flip surrounded by two domain walls. These domain walls delocalize due to the terms of the Hamiltonian that are transverse to the quantization axis, and propagate individually and independently, in contrast to the fully polarized state where they propagate in pairs as a firmly bound entity. 

The elementary excitation, the spinon, carries spin 1/2 and can be pictorially associated with an individually propagating domain wall. Spinons separate two sections of the macroscopic singlet ground state wave function that are phase-shifted by $\pi$. They are easy to visualize, Fig.~\ref{fig1}b, in the extreme Ising limit ($\Delta\!\rightarrow\!\infty$ in ${\bm S}_{n} {\bm S}_{n+1}\!=\!S_{n}^xS_{n+1}^x+S_{n}^yS_{n+1}^y+\Delta S_{n}^zS_{n+1}^z$), where they represent an abrupt domain wall between the two distinct antiferromagnetic orders. Approaching the Heisenberg limit, $\Delta\!=\!1$, the local spin flip can no longer be represented by two spinons alone, but rather decomposes in a rapidly converging series of states containing two, four, and higher even numbers of spinons.

While each spinon has a discrete energy-momentum relation, the excitation spectrum is composed of spinon pairs (and higher even-numbered spinon states with $S_{\rm tot}=1$), and will therefore appear continuous. This characteristic continuous spectrum is indeed observed at zero field, both experimentally and in exact calculations of the two- and four-spinon contributions, Fig.~\ref{fig1}d. In contrast to magnons, the spinon excitation amplitude is identical for all three orthogonal directions, $\alpha=x,y,z$. Precise calculations for the two- and four-spinon spectra (which represent around 98(1)\% of the full response function in the thermodynamic limit) have become available \cite{Caux06}, and we demonstrate in the following that our experimental results confirm the abstract spinon concept accurately on the quantitative level.

\begin{figure}[h!]
\centering
\includegraphics[width=0.8\columnwidth]{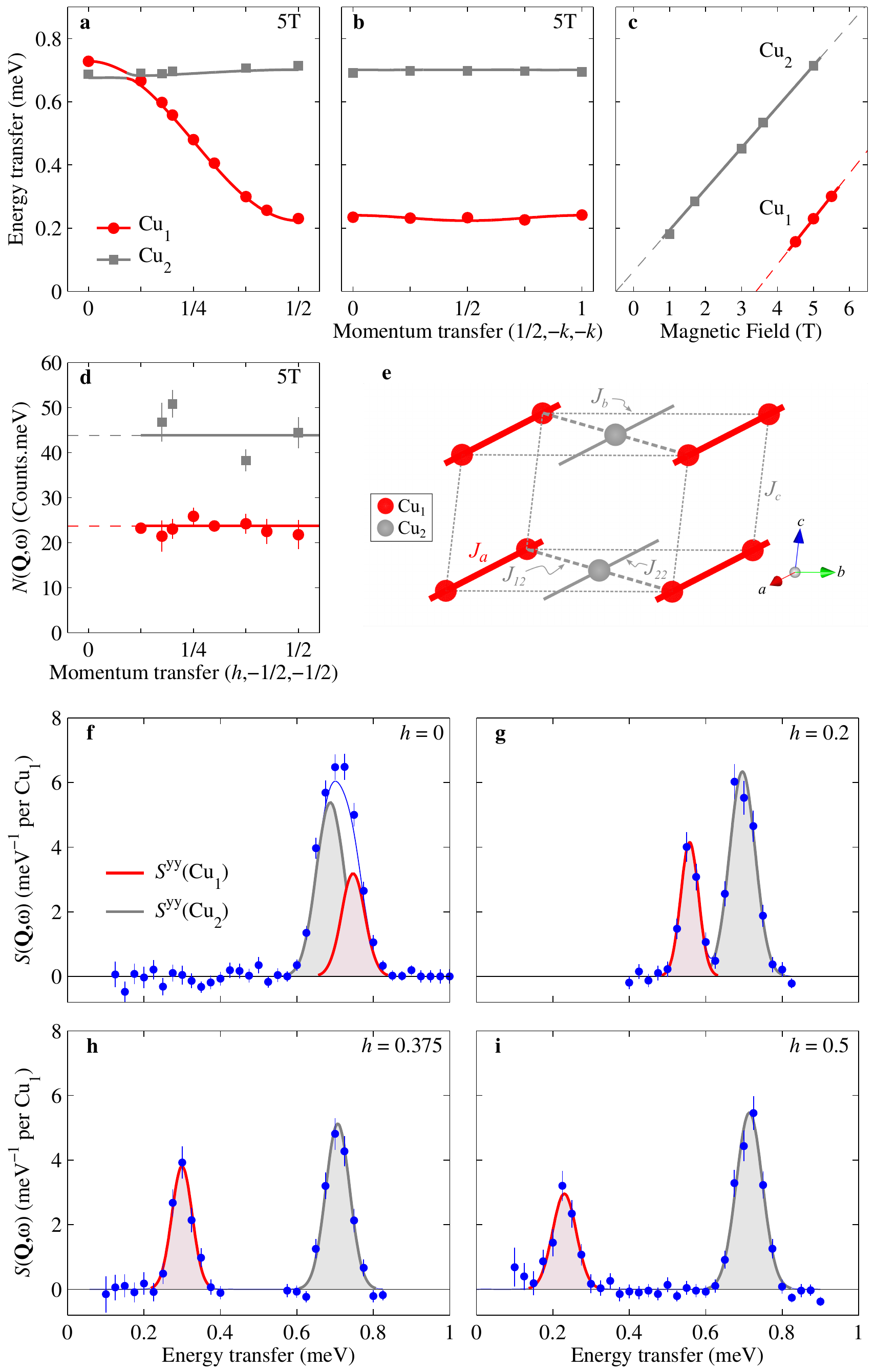}\\
\caption{}
\label{fig2}
\end{figure}

We performed inelastic neutron scattering experiments on large single crystals of \Cuso, cooled to $\sim$100~mK in a dilution refrigerator. There are two Cu$^{2+}$ sites in the elementary unit cell, Cu$_1$ at [$0,0,0$], and Cu$_2$ at [$\frac{1}{2},\frac{1}{2},0$], which each provide a localized spin 1/2, Fig.~\ref{fig2}e. All antiferromagnetic exchange is overcome by the Zeeman energy in a modest magnetic field of 5~T. In this fully polarized phase, we observe discrete excitations, well-described by a resolution-convoluted $\delta-$function, the signature of a magnon, Figs.~(\ref{fig1}c). We observe a dispersive branch with 0.517(9)~meV bandwidth along ${\bm a}^{\ast}$ and a minimum at $h=1/2$, Fig.~\ref{fig2}a. This dispersive branch reflects the principal antiferromagnetic exchange interaction between neighbouring Cu$_1$ spins which hence form chains along ${\bm a}$. Along $(0,k,k)$, perpendicular to the chain direction, the bandwidth is smaller than the experimental error of 0.007~meV, Fig.~\ref{fig2}b, and  evidences negligible coupling between individual spin chains. A second branch at energy $2g_2\mu_B H S^z$, grey in Fig.\ref{fig2}, is flat both along $(h,0,0)$ (bandwidth 0.048(6)~meV) and $(0,k,k)$ (bandwidth $<$0.007~meV). As a function of magnetic field, the dispersive branch emerges from zero energy at the saturation field strength, while the energy of the flat branch is directly proportional to the magnetic field, Fig.~\ref{fig2}. The field dependence and vanishing bandwidth of the flat branch reveal an essentially decoupled spin site (Cu$_2$), where the neutron excites the local spin from the lower to the upper Zeeman level. The Zeeman-shifts of the dispersive and of the flat branch are slightly different, as expected for the Land\'{e}-$g$-factors $g_1$ and $g_2$ of two different crystallographic sites. This scenario is also confirmed by our spin-wave calculations, which take into account various potential exchange paths sketched in Fig.~\ref{fig2}e (cf. Supplementary Materials), and provide the dominant exchange between the Cu$_1$ spins as $J_{\rm a}=0.252(17)$~meV.\\

At zero magnetic field, the Zeeman levels of the decoupled Cu$_2$ spins are not split, the flat branch has energy zero, and only the chain-forming Cu$_1$ spins contribute to the inelastic spectrum. We observe a sine-shaped lower boundary of the continuous scattering, with maximum onset-energy 0.402(2)~meV at $h=\frac{1}{4}$. This experimentally determined spinon bandwidth agrees within 2\% error with the theoretical prediction $\frac{\pi}{2}J_{\rm a}=0.406(7)$~meV, with $J_{\rm a}$ determined from the bandwidth in the fully polarized phase. We thus confirm experimentally that the energy of spinon excitations is quantum renormalized upwards by a factor $\frac{\pi}{2}$ compared to classical magnons~\cite{desCloizeaux62}.

In order to compare the observed intensities to the theoretical two- and four-spinon dynamic structure factor, we consider the inelastic neutron cross section 
\begin{equation}\label{cross}
	\frac{d^2 \sigma}{d \Omega d \omega} = \frac{k_f}{k_i}
	(\gamma r_0)^2 \hspace{-1mm}
	\displaystyle{\sum_{\alpha=x,y,z}}\hspace{-1mm}\big(g_1\,f({\bm Q})\big)^2
	\big(1-\hat{Q}^\alpha\hat{Q}^\alpha\big) 
	S^{\alpha\alpha}({\bm Q},\omega)
\end{equation}
Here, $k_i, k_f$ denote the variable incoming and fixed outgoing neutron wave vector, $\gamma$ is the gyromagnetic ratio of the neutron, $r_0$ the electron Bohr-radius, $f({\bm Q})$ the form-factor of the electronic shell responsible for the Cu$_1$ spin, taken at the total momentum transfer $\bm Q$, and $S^{\alpha\alpha}({\bm Q},\omega)=\frac{1}{(2\pi)^4\hbar}\displaystyle{\int\hspace{-2mm}\int}\, \langle S^{\alpha}(0,0)S^{\alpha}({\bm r},t)\rangle \exp\big( i ({\bm Q}\cdot{\bm r}-\omega t) \big) \,{\rm d}^3{\bm r} \,{\rm d}\,t $ the dynamic structure factor, where $\alpha=x,y,z$ denote the directions ${\bm x} || \bm Q$, ${\bm y}\perp{\bm Q}$ in the scattering plane, and $\bm z$ perpendicular to the scattering plane.

In our experiment, the magnetic field ${\bm H}||{\bm z}=[0,-1,1]$ is perpendicular to the scattering plane. The theoretical magnon intensity in the fully polarised phase is given by $S^{xx}({\bm Q},\omega)=S^{yy}({\bm Q},\omega)=\frac{S}{2}\delta\big(\omega-\omega(h)\big)$ per chain spin Cu$_1$. Of these, only $S^{yy}({\bm Q},\omega)$ is visible, since the neutron scatters exclusively from magnon eigenvectors perpendicular to the total momentum transfer ${\bm Q}||{\bm x}$ as expressed by the factor $\big(1-\hat{Q}^\alpha\hat{Q}^\alpha\big)$ in the cross section. We therefore identify the observed $k_i$-normalized intensity (cf. Supplementary Materials) with $N({\bm Q},\omega) S^{yy}({\bm Q},\omega) = N({\bm Q},\omega) \frac{S}{2}$. We observe in the experiment that $N({\bm Q},\omega)$ does not depend on ${\bm Q}$ or $\omega$, Fig.~\ref{fig2}d. $N({\bm Q},\omega)=\overline{N}$ therefore contains all prefactors in the neutron scattering cross section, including orbital and covalency effects. 


The zero-field data are then normalized by the factor $\overline{N}$ obtained from the fully saturated phase, and thus can be directly compared to the dynamic structure factor per Cu$_1$ spin. For an isotropic ground-state we have $S^{xx}({\bm Q},\omega)=S^{yy}({\bm Q},\omega)=S^{zz}({\bm Q},\omega)$, but only $S^{yy}({\bm Q},\omega)+S^{zz}({\bm Q},\omega)=2\,S^{yy}({\bm Q},\omega)$ contribute to the inelastic scattering cross section.

\begin{figure}[!h]
\centering
\includegraphics[width=0.7\columnwidth]{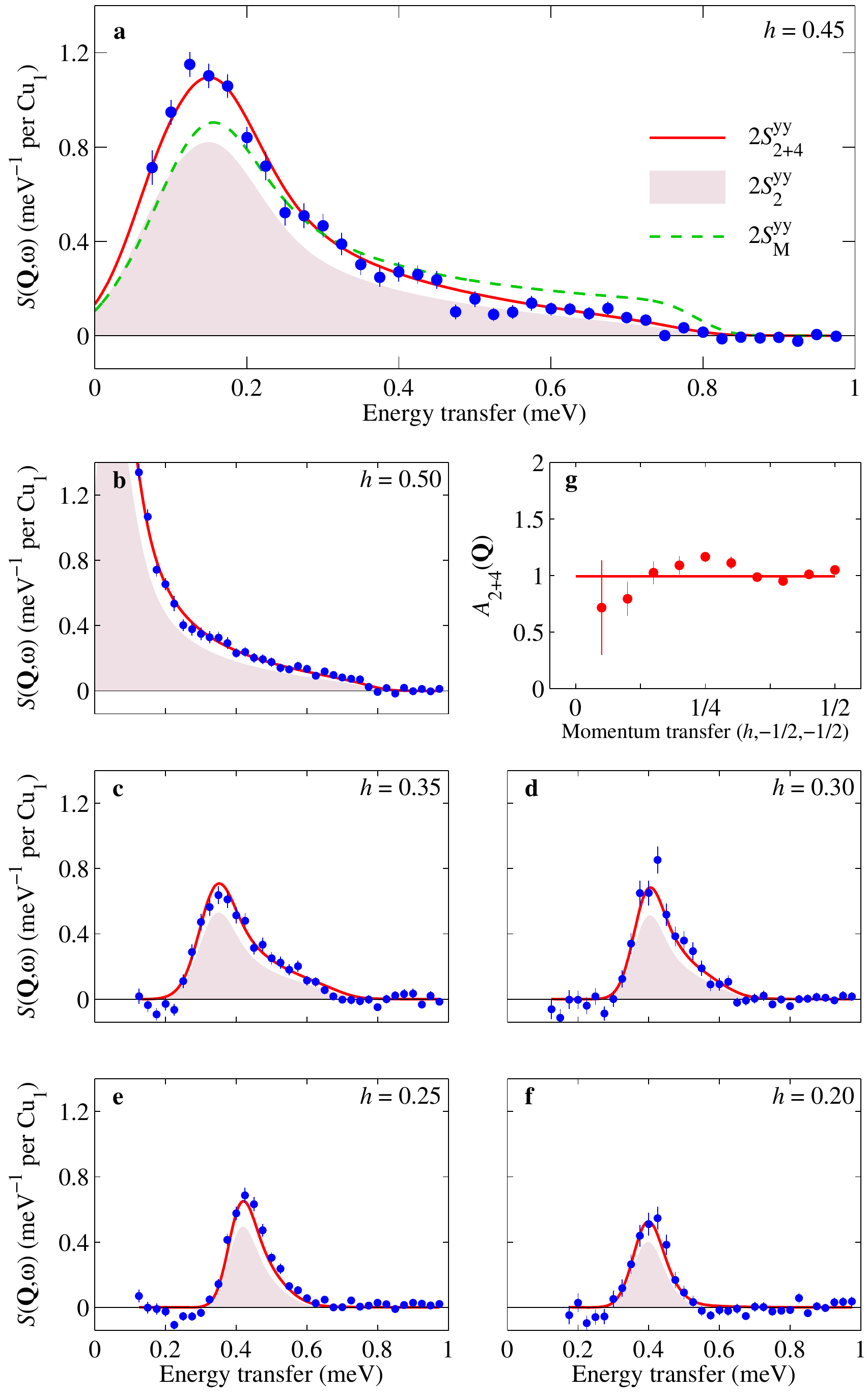}\\
\caption{}
\label{fig3}
\end{figure}

In Fig.~\ref{fig3} we compare the normalized zero-field inelastic spectra to theoretical predictions. Before exact calculations were available, experimental data were usually compared to the M\"{u}ller {\it ansatz} \cite{Mueller81}, an approximation of the two-spinon continuum with an artificial cut-off at the upper two-spinon boundary. Evidently, the lineshape of the M\"{u}ller {\it ansatz} does not very well describe our experimental data in that it underestimates the low-energy part of the spectrum and overestimates the high-energy part, Fig.~\ref{fig3}a. Next, we compare our data to the exact two- and four-spinon dynamic structure factor $2\,S^{yy}_{2+4}({\bm Q},\omega)$ \cite{Caux06}. For this comparison we introduce a ${\bm Q}$-dependent prefactor, $A_{2+4}({\bm Q})$ which equals 1 if the theory describes the normalized inelastic intensities perfectly. With the exchange $J_a=0.252$~meV fixed to the value determined by the spin-wave fit to all data of the fully polarized phase (cf. Supplementary Materials) we fit for each wave vector ${\bm Q}=(h,-\frac{1}{2},-\frac{1}{2})$ the prefactor $A_{2+4}(h)$ of the two- and four-spinon structure factor $2\,S^{yy}_{2+4}({\bm Q},\omega)$. We obtain prefactors close to 1 and essentially independent of ${\bm Q}$, with $\overline{A_{2+4}(h)}=1.03(9)$, Fig.~\ref{fig3}g. A simultaneous fit of all zero-field data with free $J_a$ yields $J_a=0.256(1)$~meV and the global prefactor $A_{2+4}=0.99(8)$. This fit is displayed as red lines in Fig.~\ref{fig3}, and is indistinguishable from lines with $A_{2+4}=1$. In order to illustrate the importance of the four-spinon contribution, Fig.~\ref{fig3}a-f also displays the exact two-spinon-only structure factor as shaded area. Inside the two-spinon boundaries, the two- and four-spinon continua have a similar lineshape. Therefore, fits to the two-spinon-only structure factor could approximately model the data, but would require an increase of the prefactor to 1.4(1). Thanks to our accurate absolute normalization of the neutron data we can therefore establish that two-spinon states only account for $74(6)\%$ of the measured spectral weight. We thereby unambiguously demonstrate that higher spinon states contribute significantly with $26(6)\%$ to the spectrum.

The essential properties of four-spinon excitations can be captured using the pictures of Fig.~\ref{fig1}b. In the Ising limit ($\Delta\rightarrow\infty$), a localized spin flip exactly projects onto a state with two spinons. Since these domain walls are localized, the state immediately after the spin flip can be represented as a combination of two-spinon states with a broad momentum distribution. In the Heisenberg case ($\Delta\rightarrow 1$) however, since each spinon is an extended object, the initial ($t=0$) state with a local spin flip must be decomposed into a quantum mechanical superposition of (mostly) two- and four-spinon states weighted by spinon-momentum dependent complex amplitudes. The evolution of that state (at a different position and later time $t>0$) is encoded in the spin-spin correlator such that two-spinon excitations contribute if one spinon propagates at the appropriate velocity. The leading four-spinon parts of the correlator closely follow that of the two-spinon states; they resemble two-spinon contributions but with two additional spinons added with close to zero momentum and energy. As a consequence, the four-spinon correlation weight is almost entirely contained within the boundaries of the two-spinon continuum, where it approximatively follows the same lineshape. In Supplementary Materials we complement this description of two-spinon and four-spinon states based on the Bethe ansatz. 

\begin{figure}[h!]
\centering
\includegraphics[width=0.98\columnwidth]{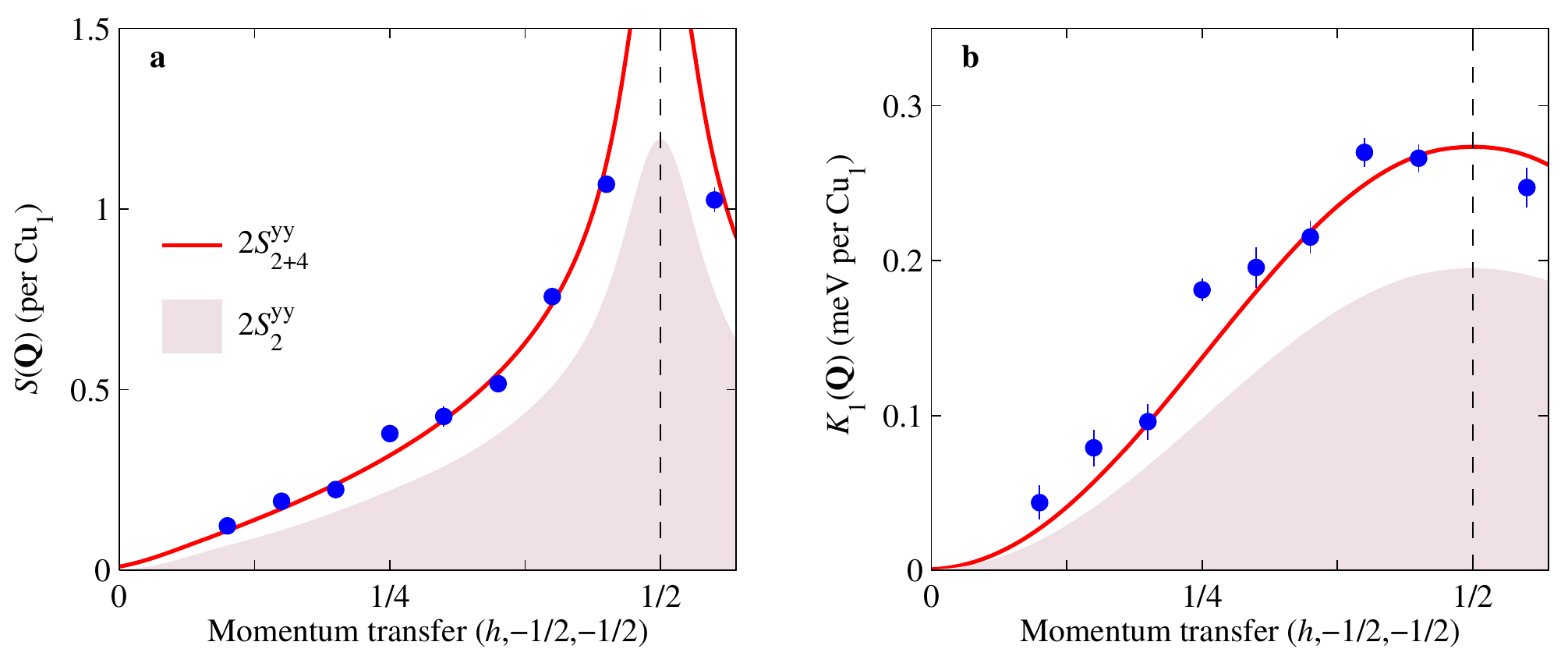}\\
\caption{}
\label{fig4}
\end{figure}

Fig.~\ref{fig4}, which illustrates the static structure factor $S({\bm Q})$ and the first frequency moment ${\int}\omega\, S({\bm Q},\omega)\, {\rm d}\omega$, additionally confirms that two- and four-spinon excitations together essentially exhaust the spectral weight and the first moment sum rule. The two-spinon-only contribution is again displayed as shaded area, and can clearly not account for the observed intensity. Our work thus proves quantitatively the validity of the spinon concept for the excitation spectrum of the spin 1/2 Heisenberg chain.


In conclusion, we exploit in this work that a large magnetic field quenches \Cuso\ from a macroscopically entangled quantum state into the fully polarized state that can be described classically. Domain wall pair excitations have then a finite threshold energy and are bound or ``confined'' by the magnetic field. The resulting quasiparticle, the magnon, is correctly described by the classical spin wave theory. We use its known energy and intensity to quantitatively confirm the abstract concept of fractional spinon excitations out of the macroscopically entangled quantum state at zero-field in a real material. The theoretical spinon concept is fully confirmed by our experiment - the spinon has a $\pi/2$ larger bandwidth than the magnon, the lineshape of the energy scans corresponds to the exact two-and four spinon dynamic structure factor, and both two- and four spinon contributions are needed to account for the experimentally observed intensity, spectral weight, static structure factor, and first moment sum rule.

\NL
{\bf Methods}
\NL
\NI Full Methods are available in Supplementary Materials.

Inelastic neutron scattering experiments were performed on the cold triple axis spectrometer IN14 at the Institut Laue-Langevin (Grenoble), equipped with PG(002)-monochromator and analyser and final neutron wave vector fixed at $k_f=1.15$\AA$^{-1}$ (FWHM energy resolution of 0.078~meV) or 1.025\AA$^{-1}$ (FWHM energy resolution of 0.05~meV). The sample was a 2~cm$^3$ single crystal of deuterated \Cuso\ oriented with $(h,0,0)$ and $(0,k,k)$ in the horizontal scattering plane, and cooled to 100~mK (above its N\'{e}el ordering temperature) by a dilution insert inside either an ILL-orange type cryostat or a 5.5~T vertical magnet. The non-magnetic background from incoherent elastic scattering was derived from the high-field measurements and subtracted from all spectra.

The spectrum from the high-field fully polarized state was analysed by a linear spin-wave theory fit to yield exchange parameters, $g-$factors, and absolute intensity. We obtained a dominant exchange between nearest-neighbour Cu$_1$ spins, $J_{\rm a}=0.252(17)$~meV. Other magnetic interactions in the material were found to be small or negligible, $J_{\rm b}+J_{\rm c}<$0.004(7)~meV, $J_{22}=-0.012(18)$~meV and $J_{12}=-0.020(22)$~meV. The $g-$factors for the magnetic field along [$0,1,-1$] were determined for each copper site from the Zeeman shift of their respective branch. The theoretical magnon intensity was obtained in the fully polarized phase as $S^{xx}({\bm q},\omega)=S^{yy}({\bm q},\omega)=\frac{S}{2}\delta(\omega-\omega(h))$ per Cu$_1$.

The zero field data were analysed by comparing to calculations including two- and four-spinon states. Their exact contributions to the zero-temperature dynamical structure factor of the spin 1/2 Heisenberg isotropic chain were obtained directly in the thermodynamic limit using the vertex operator approach~\cite{Jimbo95} based on the exact solvability of the Heisenberg model~\cite{Caux06}. This theoretical dynamic structure factor was compared to the zero-field data after convolution to a normalized two-dimensional Gaussian profile to account for the finite experimental energy and momentum resolution in the experiment. From a global least-squares fit to the entire dataset we obtained $J_a=0.256(1)$~meV, a global amplitude $A_{2+4}=0.993(84)$, and an energy resolution of 0.078~meV, in perfect agreement with the measured resolution at zero energy transfer.

\NL

\clearpage

\NL
{\bf Acknowledgments}
\NL
We acknowledge useful discussions with C.~Broholm, B.~Dalla Piazza, B.~F\aa{}k, B. Lake, C.~R\"uegg, and A. Tennant. The work of M.M. was supported in part by U.~S. Department of Energy (DOE), Office of Basic Energy Sciences, Division of Materials Sciences and Engineering under award DE-FG02-08ER46544. M.~E. acknowledges support from the Deutsche Bundesministerium f\"ur Bildung, Wissenschaft, Forschung und Technologie (BMBF), project 03KN5SAA. H.~M.~R. acknowledges support from the Swiss National Science Foundation (SNF) and the European Research Council (ERC). J.-S.~C. acknowledges support from the Foundation for Fundamental Research on Matter (FOM) and the Netherlands Organisation for Scientific Research (NWO).

\NL
{\bf Author contributions}
\NL
M.~E. and H.~M.~R. performed the experiment with the help of A.~S. on a crystal synthesized by A.~K. Data treatment and fits were carried out by M.~M., M.~E. and H.~M.~R. Exact theoretical calculations were performed by J.-S.~C. The physical pictures for multi-spinon excitations were developed through various discussions between J.-S.~C. and M.~M., M.~E. and H.~M.~R; M.~E., M.~M., J.-S.~C and H.~M.~R wrote the manuscript.   

\NL
{\bf Competing financial interests}
\NL
The authors declare no competing financial interests.

\clearpage
{\bf Figure {1}. Cartoon representation of the magnetic excitations in a spin-1/2 (Heisenberg) antiferromagnetic chain and overview of the neutron scattering results for \Cuso.} ({\bf a})Fully polarized (saturated) state. The creation of a magnon by inelastic scattering of a neutron can be imagined as a single spin flip. The Zeeman energy prevents any growth of the flipped section that propagates like a single entity. This magnon can classically be visualized as a spin wave, a coherent precession of the local spin expectation value around the field direction. ({\bf b}) Zero-magnetic field state. Snapshots of large antiferromagnetically correlated regions of the ground state. The spins could be found in a locally antiferromagnetic configuration with equal probability in any direction ({\it e.g.}\ the opposite one (shadows)). The neutron acts on the singlet ground state and excites triplet states which we may imagine as a local spin-flip surrounded by two domain walls, which individually correspond to a spinon carrying spin 1/2. The spatial extent of a spinon depends on the anisotropy: in the Ising limit, a local spin-flip decomposes into two spinons; in the Heisenberg limit, it decomposes into a rapidly converging series of states containing two, four, and higher even numbers of such spinons. ({\bf c}) Intensity maps of the experimental and theoretical magnon dispersion in the fully polarized phase of \Cuso\ for $\mu_0 H=5$~T$>\mu_0H_{\rm sat}$. The cosine-shaped dispersion corresponds to the excited magnon of the saturated Heisenberg antiferromagnetic chain and the additional flat branch around 0.7~meV is a transition between two local Zeeman-levels of a second decoupled Cu-site in \Cuso.  ({\bf d}) Intensity colourmaps of the experimental inelastic neutron scattering  spectrum of the chain spins in the zero-field phase of \Cuso, and theoretical two- and four-spinon dynamic structure factor.
\clearpage

\clearpage
{\bf Figure 2. Excitations in the fully polarized state ($\mu_0 H=5$~T).} ({\bf a}) Dispersion and ({\bf d}) intensity of the Cu$_1$ antiferromagnetic chain (red circles) and the  flat transition between local Zeeman-states of Cu$_2$ (grey squares). The lines display the spin wave fit described in Supplementary Materials. ({\bf b}) Flat dispersions along $(0,k,k)$ evidence that Cu$_1$ and Cu$_2$ spins are decoupled in this direction. The lines represent the spin wave fit. ({\bf c}) The Zeeman-shift of Cu$_1$ and Cu$_2$ branches at the wave-vector $(\frac{1}{2},-\frac{1}{2},-\frac{1}{2})$ extrapolates to quasi-zero field for the decoupled Cu$_2$, and to the saturation field for the Cu$_1$ chain magnon. The straight line fits correspond to $g_1$ and $g_2$ given in the text. ({\bf e}) Magnetic primitive cell, triclinic symmetry. Chain-forming Cu$_1$ sites (red) at [$0,0,0$] and decoupled Cu$_2$ sites (grey) at [$\frac{1}{2},\frac{1}{2},0$]. ({\bf f}-{\bf i}) Selected energy scans at constant wave vector $(h,-\frac{1}{2},-\frac{1}{2})$ together with spin wave fit, described in Supplementary Materials.\\

\clearpage
{\bf Figure 3.  Excitations in zero magnetic field.} ({\bf a}-{\bf f}) Experimentally determined $S(h,\omega)=S^{yy}(h,\omega)+S^{zz}(h,\omega)$ (blue circles) in comparison to the two- and four-spinon structure factor $2\,S^{yy}_{2+4}(h,\omega)$ (red line), and the two-spinon only intensity $2\,S^{yy}_{2}(h,\omega)$ (rose shaded area), for typical wave-vectors ${\bm Q}=(h,-\frac{1}{2},-\frac{1}{2})$ with $h$ as indicated in the legend. The theoretical structure factors are shown for $J_{\rm a}=0.256$~meV, obtained from a global fit. In ({\bf a}), the experimental data are compared not only to the exact two- and four-spinon structure factor (red line) but also to a fit to the M\"{u}ller {\it ansatz} \cite{Mueller81} $2\,A_M S^{yy}_{M}(h,\omega)$ (green dashed line) with free amplitude $A_M$, cf. Text and Supplementary Materials. ({\bf g}) Fitted amplitude $A_{2+4}(h)$ for $2\,S^{yy}_{2+4}(h,\omega)$, with $J_{\rm a}=0.252$~meV fixed as determined from the spin wave fit in the fully polarized phase (red circles). The red line corresponds to $A_{2+4}=0.99$ obtained from the same global fit as for the other panels.

\clearpage
{\bf Figure 4. Sum Rules.} ({\bf a}) Experimental structure factor $S({\bf Q})={\int} S({\bf Q},\omega){\rm d}\omega={\int}\big(S^{yy}({\bf Q},\omega)+S^{zz}({\bf Q},\omega)\big){\rm d}\omega$ (blue circles) and ({\bf b}) first frequency moment  $K_1({\bf Q})={\int}\omega\, S({\bf Q},\omega)\, {\rm d}\omega$ (blue circles) as a function of momentum transfer ${\bf Q}=(h,-\frac{1}{2},-\frac{1}{2})$ versus two- and four-spinon calculations, $2\,S_{2+4}^{yy}({\bf Q},\omega)$ (red line) and $2\,S_{2}^{yy}({\bf Q},\omega)$ (shaded area).
\clearpage

{\noindent  \large \bf Supplementary Materials for \textit{Fractional spinon excitations in the quantum Heisenberg antiferromagnetic chain}.}

\NL
{\bf Inelastic neutron scattering}
\NL
We performed inelastic neutron scattering experiments on the cold triple axis spectrometer IN14 at the Institut Laue-Langevin (Grenoble), equipped with PG(002)-monochromator and analyser, and collimation 40'-40'-40'-open, keeping the final neutron wave vector fixed at $k_f=1.15$\AA$^{-1}$ (($h,0,0$)-dispersion). We also used 1.025\AA$^{-1}$ (for the $(0,k,k)$-dispersion at $\mu_0H=5$~T and part of the Zeeman-shifts), with the analyser horizontally focused. A Beryllium-filter, cooled to 77~K was used in $k_i$. The measured energy resolution (full width at half maximum) was 0.078~meV  for $k_f=1.15$\AA$^{-1}$ and 0.05~meV for $k_f=1.025$\AA$^{-1}$. All data were normalized to the flux-monitor, installed in $k_i$. Since the intrinsic sensitivity of this flux-monitor is proportional to $k_i^{-1}$, the $k_i$-dependence of the cross section is accounted for by the normalization of all data to the monitor. 

The sample, a 2~cm$^3$ single crystal of deuterated \Cuso, similar to that pictured in Fig.~{s1}, was oriented with $(h,0,0)$ and $(0,k,k)$ in the horizontal scattering plane, and cooled to 100~mK by a dilution insert inside either an ILL-orange type cryostat or a 5.5~T vertical magnet. All data were taken above the N\'{e}el ordering temperature into three-dimensional long-range order. Normalization between the data taken with the magnet and with the orange cryostat was obtained via an energy scan taken at zero field and 100~mK on the same crystal in both environments. 
\begin{center}
\includegraphics[width=0.40\columnwidth]{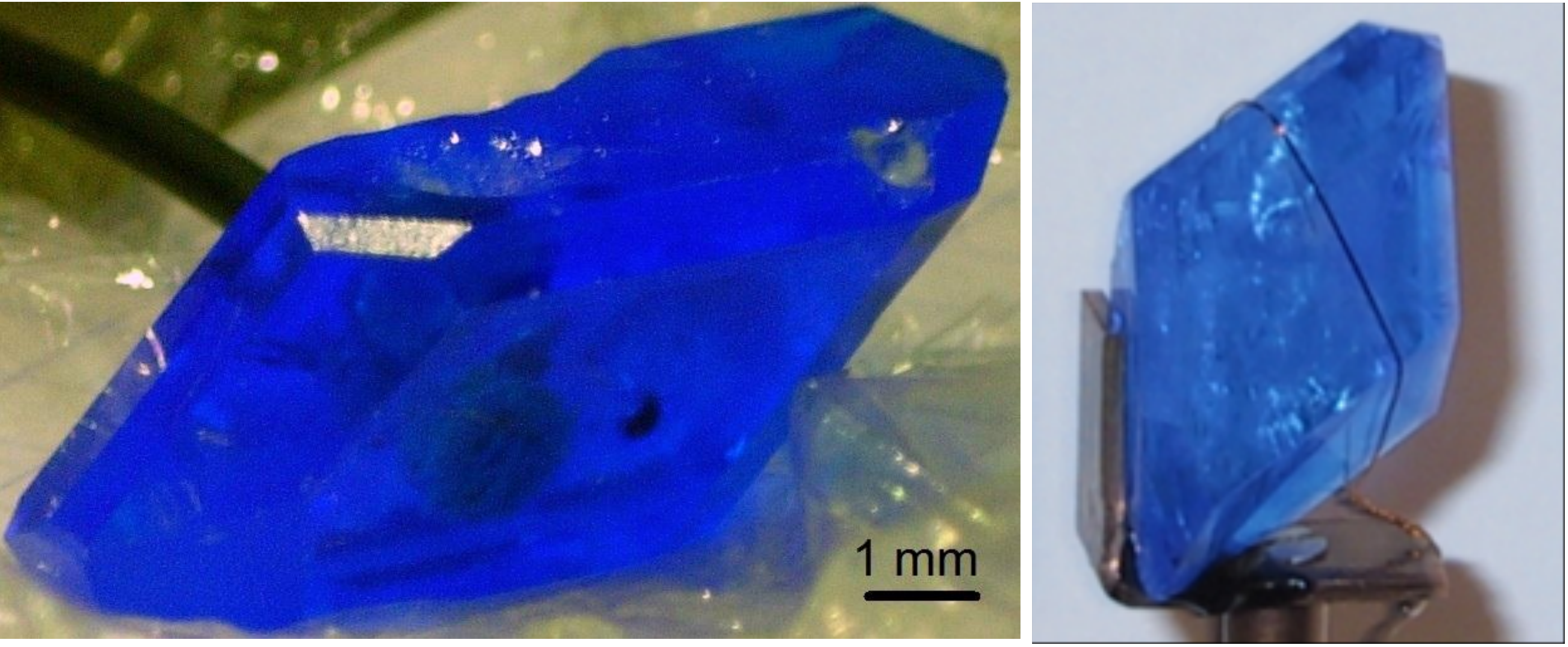}\\
Fig. {\bf s1} Single crystals of deuterated \Cuso.
\end{center}

At 5~T and 5.5~T, all magnetic excitations are sharp and driven to high energies. The non-magnetic background, arising from substantial incoherent elastic scattering (probably due to H-D exchange of the sample's crystal water when exposed to air), can therefore be easily subtracted from all spectra. In order to isolate the signal coming from Cu$_1$ at zero magnetic field, the amplitude of the incoherent scattering (including the paramagnetic scattering from Cu$_2$) can be determined at energy transfer zero for each wave vector, since even at $h=\frac{1}{2}$, the magnetic contribution from Cu$_1$ is only a few percent ($<$4\%) of the incoherent  scattering. The energy-profile of the incoherent scattering at non-zero energies can be measured at $h=0$ where the scattering from Cu$_1$ spins is zero for all energy transfers (the structure factor of the Heisenberg antiferromagnetic chain is zero at $q=0$), and above the upper continuum boundary for various wave vectors. Everywhere else the background is obtained by interpolation and then subtracted from the data. 

\NL
{\bf Spin-wave theory}
\NL
A spin-wave fit of all data in the fully polarized phase shows that the exchange between nearest-neighbour Cu$_1$ spins along the ${\bm a}$-axis, $J_{\rm a}=0.252(17)$~meV, is responsible for the bandwidth $2{J}_a$ of the dispersive magnon, while the tiny dispersion bandwidth perpendicular to ${\bm a}^{\ast}$ limits $J_{\rm b}+J_{\rm c}$ to 0.004(7)~meV. The interactions linking the Cu$_2$ spins are found negligible, it couples ferromagnetically along ${\bm a}$ with $J_{22}=-0.012(18)$~meV, and to the first site Cu$_1$ with $J_{12}=-0.020(22)$~meV. The $g-$factor for the magnetic field along [$0,1,-1$] is determined from the Zeeman shift of the respective branch. We find $g_1=2.49(6)$, $g_2=2.25(5)$. These agree well with the findings for Cu$_1$ and Cu$_2$ from EPR on hydrogenated \CusoH\ at room temperature \cite{Bissey95}. Since \Cuso\ has only one chain-spin Cu$_1$ per crystallographic unit cell, the applied magnetic field does not induce a staggered local magnetic field onto the spins. 
\NL

The theoretical magnon intensity is given by $S^{xx}({\bm q},\omega)=S^{yy}({\bm q},\omega)=\frac{S}{2}\delta(\omega-\omega(h))$ per Cu$_1$. With $N$ spins coupled to a HAF-chain via $J_{\rm a}$, there are $N$ wave vector states per Brillouin zone, each contributing with $S$ to the total spectral weight. The remaining spectral weight, $N S^2$, resides at $\omega=0, {\bm q}=0$ in the ferromagnetic Bragg peak of the fully polarized chain, so that chain-magnon and ferromagnetic Bragg peak together exhaust the entire spectral weight $NS(S+1)$ or $S(S+1)$ per chain spin Cu$_1$. In experiments with a finite resolution the delta-function needs to be replaced by a normalized Gaussian. For the small ${\bf Q}$-range and temperature (100~mK) of our experiments, the Debye-Waller factor is well approximated by 1.

\NL

In Fig.~\ref{fig2} the intensity of the flat branch appears substantially higher than that of the dispersive branch. This could either point to additional free Cu-spins (due to imperfections of the crystal) and/or result from multiple scattering of the neutron (an incoherent elastic scattering process combined with scattering from the flat excitation branch). In view of the size of the crystal and the amplitude of the incoherent elastic scattering, the latter appears much more plausible. For dispersive excitations, multiple scattering would affect the intensities at the maximum and minimum of the dispersion band much more than the intensity in the middle of the band. Since we do not see any visible increase of $N({\bf Q,\omega})$ at $h=\frac{1}{2}$ compared to $h=\frac{1}{4}$ we conclude that multiple scattering from dispersive excitations is still negligible. This means that we can neglect it for the evaluation of the zero-field intensities.

\NL
{\bf Zero-field model}
\NL
The zero-field data are compared to the exact contributions from two-spinon and four-spinon states to the zero-temperature dynamical structure factor of the spin 1/2 Heisenberg isotropic chain, obtained directly in the thermodynamic limit using the vertex operator approach~\cite{Jimbo95} based on the exact solvability of the Heisenberg model~\cite{Caux06}. This theoretical dynamic structure factor was convoluted to a normalized two-dimensional Gaussian profile to account for the finite experimental energy and momentum resolution. A global least-squares fit to the data was performed with four free parameters: the chain exchange $J_{\rm a}$, the global amplitude prefactor $A_{2+4}$, and the energy and momentum resolution. We obtain a unique set of parameters that describes the entire dataset, as shown in Fig.~\ref{fig2}. This fit yields $J_a=0.256(1)$~meV, and $A_{2+4}=0.993(84)$, and an energy resolution (full width at half maximum) of 0.078~meV, in perfect agreement with the measured resolution at zero energy transfer (see above). In a second step, the resolution was kept fixed, the exchange constant $J_{\rm a}$ was kept fixed to the value determined from the spin-wave fit above the saturation field, $J_{\rm a}=0.252$~meV, while the amplitude prefactor was allowed to change for each momentum-transfer. This second fit gives consistent results, see the $A_{2+4}(h)$ in Fig.~\ref{fig2}g, with an average $\overline{A_{2+4}(h)}=1.03(9)$. Since $k_i$ varies only by 20\% over the whole range of energy transfers, we do not expect a visible change of the resolution with energy. 
\NL

The experimental structure factor $S({\bf Q})={\int}S({\bf Q},\omega){\rm d}\omega$ and the experimental first moment $K_1({\bf Q})={\int}\omega\, S({\bf Q},\omega)\, {\rm d}\omega$ are obtained by numerical integration of the constant-${\bf Q}$ scans using data above 0.150~meV. Close to $h=0.5$, where the lower continuum boundary approaches zero energy transfer, we naturally miss out some intensity.

\NL
{\bf Additional physical picture for two- and four-spinon states}
\NL
An alternative picture for multi-spinon excitations in the spin 1/2 HAF chain, at any anisotropy $\Delta$, can be obtained in the Bethe ansatz language, in which the zero-field ground state is described by a distribution of filled quantum numbers. Spinons are holes within this distribution and can be seen as somewhat analogous to holes for a filled Fermi sea. However, in contrast to fermions, spinons are semions~\cite{Haldane91} and adding two of them only blocks \textit{one} available quantum number. Low-energy spinons are associated with holes that occupy one or several quantum numbers lying close below the highest occupied quantum numbers defining the Bethe ansatz ground state. The latter quantum numbers, designated as ``Fermi Level'' in the following, are those for which creating holes results in spinons with zero energy. As holes are created deeper into the sea, the energy of the spinons increases. In the XY limit, $\Delta \rightarrow 0$, the (longitudinal) spin-spin correlator is completely exhausted by two-spinon states. As they all have the same correlation weight, the dynamic structure factor simply follows the two-spinon density of states. Increasing the anisotropy, $\Delta \rightarrow 1$, has two effects: the two-spinon matrix elements become momentum-dependent, and higher-spinon states obtain correlation weight~\cite{Caux11}. Of all two-spinon states, those associated with holes that stay close to the Fermi level contribute most to the dynamic structure factor. The lower threshold of the excitation continuum (where the correlator is singular) is formed by states having one hole at the Fermi level and the other dispersing through the sea. Similarly, of all four-spinon states, the most important contributors are those associated with one dispersing hole and three additional holes situated close to the Fermi level. This implies that around the lower boundary of the continuum, the participation of four-spinon states to the dynamical structure factor closely follows that of two-spinon states. Although four-spinon contributions are not confined to the two-spinon continuum, they only carry sizeable correlation weight within the boundaries of the latter.

\end{document}